\documentclass[aps,prl,twocolumn,groupedaddress]{revtex4-2}

\usepackage{amsmath}
\usepackage{amssymb}
\usepackage{graphicx}
\usepackage{bm}          
\usepackage{hyperref}    

\begin{document}


\title{Parametric Decay of Broadband Lower Hybrid Waves}


\author{S. Jin, A. K. Ram, S. G. Baek, P. T. Bonoli}
\affiliation{MIT Plasma Science and Fusion Center, Cambridge, MA 02139, USA}


\date{\today}

\begin{abstract}
We derive a nonlinear dispersion relation for the parametric decay of an arbitrary lower hybrid pump spectrum, generalizing the standard three-wave treatment to the broadband case. Finite pump bandwidth is found to reorganize the decay spectrum, collapsing the nonresonant continuum predicted by monochromatic theory into narrow bands localized at ion cyclotron harmonics. This reorganization is distinct from the gradual suppression usually associated with finite pump coherence: it occurs at modest bandwidth, while growth rates remain large. Accounting for finite bandwidth restores agreement between theoretical growth rate spectra and sideband measurements from lower hybrid current drive experiments on Alcator C-Mod. More broadly, these results suggest that pump coherence can qualitatively determine the accessible decay channels in magnetized plasmas with kinetic resonances.\end{abstract}


\maketitle

Parametric decay instabilities (PDIs) are ubiquitous in plasma physics, arising whenever a wave of sufficient amplitude drives coupling to secondary excitations. In tokamaks, they feature prominently during lower hybrid current drive (LHCD)\cite{Baek2013,Li2024,Dyachenko2019,Cesario1989}, a primary tool for non-inductive current drive and a leading candidate for sustaining steady-state reactor operation \cite{Fisch1987,Bonoli2014}. In LH PDI, the launched pump wave decays into a lower hybrid sideband and a low-frequency ion quasimode. As this process modifies the launched spectrum in the edge region and becomes increasingly pronounced at higher plasma density, it is suspected to underlie the  anomalous degradation of current drive efficiency observed well below the classical accessibility limit\cite{Liu1984,Wallace2010,Li2022}. Accurate modeling of LH PDIs is therefore essential for predicting LHCD performance, especially at the higher densities relevant for future reactor scenarios.

The lower hybrid regime is a particularly demanding test of PDI theory: strong magnetization and kinetic ion effects give rise to competing decay channels whose characteristic sideband spectra are directly accessible to experiment. Although standard monochromatic theory \cite{Porkolab1977,Liu1986} captures the existence of these distinct decay channels, their predicted relative dominance can be at odds with experimentally observed spectra. For example, at the onset of PDI activity, sideband spectra measured on Alcator C-Mod exhibit sharp peaks separated by multiples of the local ion cyclotron frequency---the signature of ion cyclotron quasimode (ICQM) decay \cite{Baek2014, Baek2015}. Yet for the same conditions, monochromatic theory predicts that this discrete structure should be drowned out by broadband nonresonant quasimode (NRQM) activity filling the continuum between cyclotron harmonics. 

One aspect that has not been treated is the finite bandwidth of the LH pump spectrum---launchers produce waves with a spread in wavenumber \cite{Brambilla1976}, while turbulent fluctuations in the plasma edge broaden the spectrum further \cite{Biswas2022}. The role of pump bandwidth has attracted considerable attention in other contexts, notably intertial confinement fusion, as finite bandwidth can strongly suppress parametric growth \cite{Thomson1974,Bates2023}. Phase-space methods based on the Wigner-Moyal formalism \cite{Tracy2014} have recently been applied to model such broadband processes \cite{Santos2007,Brando2021,Ruskov2024}. These treatments, however, have been restricted to unmagnetized plasmas, described by relatively simple fluid equations---whereas magnetization and kinetic ion effects are essential to LH PDIs.

Here we present the first broadband treatment of LH PDIs, deriving a nonlinear dispersion relation (NLDR) for parametric decay driven by an arbitrary pump spectrum. We find that finite pump bandwidth reorganizes the decay spectrum, collapsing the NRQM continuum predicted by monochromatic theory into narrow bands localized at ion cyclotron harmonics. Crucially, this reorganization occurs before the largest ICQM growth rates are substantially suppressed, and as such, is a distinct phenomenon from the usual gradual suppression of PDIs via finite pump coherence. Applied to a well-diagnosed LH PDI event on Alcator C-Mod, the broadband treatment recovers the structure of the measured sideband spectrum. Our results establish pump coherence as a qualitative determinant of the LH PDI decay spectrum.

We use an electrostatic hybrid fluid-kinetic framework in which the nonlinear electron dynamics are treated via a drift-reduced isothermal fluid model and ions are treated kinetically but linearly. Such a hybrid approach captures the essential physics of LH PDIs while remaining analytically tractable, and has been validated against fully kinetic and electromagnetic treatments \cite{Gao2025,Liu2020}. The base equations are:
\begin{gather}
    \partial_t n + \nabla \cdot(n\boldsymbol{v}) = 0, \label{eq:n}\\
    [\partial_t + (\boldsymbol{v}\cdot\nabla)]u = -v_T^2 \partial_z(\phi + \ln n), \label{eq:u}\\
    -l_D^2 \nabla \cdot(1 + \hat{\chi}_i) \cdot \nabla\phi = n\label{eq:p},
\end{gather}
where $\boldsymbol{v} = (v_T^2/\omega_{ce})\hat{\boldsymbol{z}}\times\nabla\phi + u\hat{\boldsymbol{z}}$
is comprised of the $\boldsymbol{E}\times\boldsymbol{B}$ drift and the parallel electron
velocity $u$, with $\boldsymbol{B}_0 = B_0\hat{\boldsymbol{z}}$,
$\phi \doteq q_e\varphi/T_e$ is the normalized electrostatic potential, $n$ is the
electron density normalized to the equilibrium density $n_0$, $v_T = \sqrt{T_e/m_e}$, $l_D = v_T/\omega_{pe}$,
$\omega_{ce} = q_eB_0/m_ec$, and $\hat{\chi}_i$ is the kinetic ion
susceptibility \cite{Stix}.

For LH PDIs, the pump frequency $\omega_0$ is well separated from 
the decay mode frequency $\Omega \sim \omega_{ci} = q_iB_0/m_ic$, motivating a decomposition into 
slow components $(\bar{n}, \bar{u}, \bar{\phi})$ describing the fields associated 
with the decay mode and fast components $(\widetilde{n}, \widetilde{u}, \widetilde{\phi})$ 
describing the pump and sideband fields, where overbars denote averages over 
timescales long compared to $\omega_0^{-1}$ but short compared to $\omega_{ci}^{-1}$. Substituting this decomposition into Eqs.~\eqref{eq:n}--\eqref{eq:p} and averaging yields evolution equations for the slow fields 
\begin{gather}
    \partial_t \bar{n} + \partial_z \bar{u} = S_n,\label{eq:dens} \\
    \partial_t \bar{u} + v_T^2\partial_z(\bar{n} + \bar{\phi}) = S_u, \\
    -l_D^2\nabla\cdot(1+\hat{\chi}_i)\cdot\nabla\bar{\phi} = \bar{n}\label{eq:slowp},
    \end{gather}
    where the source terms $S_n$ and $S_u$ are slow averages of terms bilinear in the fast fields; $S_u$ is the 
broadband generalization of the usual ponderomotive force, while $S_n$ represents a 
nonlinear density source that is typically neglected in monochromatic treatments.

The source terms are fully determined by two-point statistics of the fast fields, which are naturally encoded in the averaged Wigner matrix $\boldsymbol{W}$ with components \begin{gather}
    W_{\alpha\beta}(t,\boldsymbol{x},\boldsymbol{k}) 
= \int d\boldsymbol{s}\, e^{-i\boldsymbol{k}\cdot\boldsymbol{s}}\,\overline{\psi_\alpha(t,\boldsymbol{x}_+)\psi_\beta^*(t,\boldsymbol{x}_-)},
\end{gather} 
where $\boldsymbol{x}_\pm\doteq\boldsymbol{x}\pm\boldsymbol{s}/2$ and $\boldsymbol{\psi} = (\tilde{n}, \tilde{u})^T$. Its evolution follows from the fast-field equations within the quasilinear approximation---the broadband generalization of the standard three-wave truncation---yielding the Wigner-Moyal equation \cite{Tracy2014}:
\begin{gather}
    i\partial_t \boldsymbol{W} = \boldsymbol{H}\star \boldsymbol{W} - \boldsymbol{W}\star \boldsymbol{H}^\dagger,\label{eq:WME}
\end{gather}
where $\star$ denotes the Moyal product, ($A \star B \doteq A\,e^{i\hat{\mathcal{L}}/2}B$, with 
$\hat{\mathcal{L}} = \overleftarrow{\partial}_{\bm{x}}\cdot\overrightarrow{\partial}_{\bm{k}} 
- \overleftarrow{\partial}_{\bm{k}}\cdot\overrightarrow{\partial}_{\bm{x}}$) \cite{Tracy2014}.
The Hamiltonian $\boldsymbol{H} = \boldsymbol{H}_0(\boldsymbol{k}) + 
\boldsymbol{h}(t,\boldsymbol{x},\boldsymbol{k};\bar{n},\bar{u},\bar{\phi})$ is composed of $\boldsymbol{H}_0$ 
encoding the free wave dynamics (its eigenvalues $\pm\omega(\bm{k})$ satisfy the electrostatic electron plasma wave dispersion relation
$\omega^2(\bm{k}) = (v_T^2+\omega_{pe}^2/k^2)k_z^2$), and an inhomogeneous part $\boldsymbol{h}$ encoding 
coupling to the decay fields.  At fast timescales $\omega \sim \omega_0$, the ion 
response is negligible and the fast potential is determined by the fast density 
via $-l_D^2\nabla^2\tilde{\phi} = \tilde{n}$.  Equations~\eqref{eq:dens}-\eqref{eq:WME} form a reduced system governing the coupled evolution of the slow decay fields and the fast-field statistics encoded in $\boldsymbol{W}$; their derivation and explicit forms of the longer expressions therein are given in SM.
\begin{figure}[t]
    \includegraphics[width=\columnwidth]{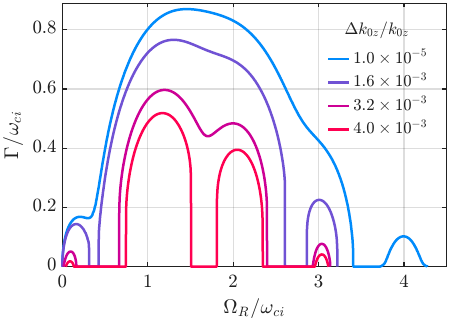}
    \caption{Maximum growth rate envelope $\Gamma(\Omega_R) \doteq \max\{\mathrm{Im}\,\Omega 
    : \mathrm{Re}\,\Omega = \Omega_R\}$ for the perpendicular coupling limit 
    ($\theta = \pi/2$), for plasma and pump wave parameters matched to the LFS LCFS conditions of 
    Ref.~\cite{Baek2014}, and increasing parallel pump bandwidth $\Delta k_{0z}/k_{0z}$ 
    (blue to red).}
    \label{fig:spectrum}
\end{figure}

\begin{figure*}[t]
    \includegraphics[width=\textwidth]{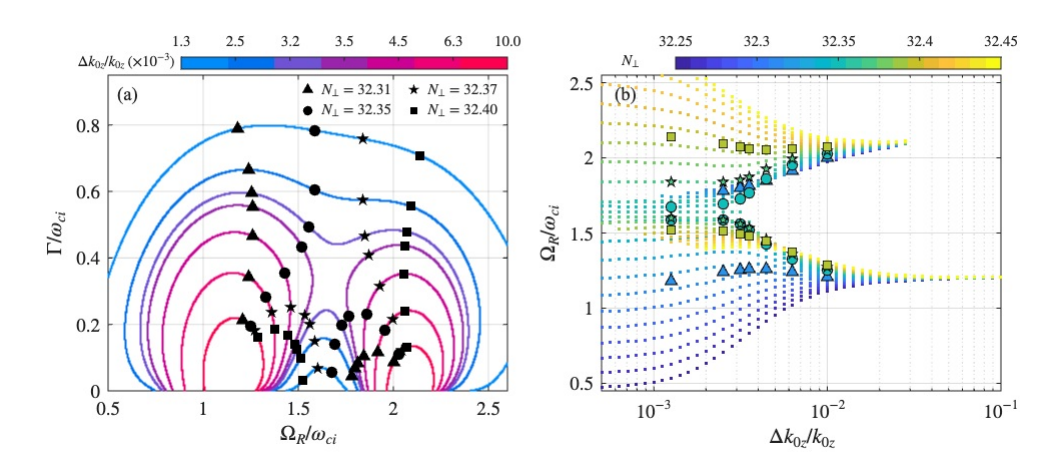}
    \caption{Reorganization of the decay mode spectrum with increasing pump bandwidth
    $\Delta k_{0z}/k_{0z}$, for same parameters as figure 1. (a) All solutions of the broadband NLDR~(\ref{eq:nldr}) 
    in the $(\Omega_R, \Gamma)$ plane at select bandwidth values, with four 
    representative values of $N_\perp$ tracked across bandwidth values (markers). (b) Real frequency $\Omega_R/\omega_{ci}$ as a function of 
    $\Delta k_{0z}/k_{0z}$ for a dense set of $N_\perp\doteq c K_\perp/\omega_0$ values (colored by $N_\perp$). The markers indicate the same $N_\perp$ values tracked in (a).}
    \label{fig:reorg}
\end{figure*}
To derive the NLDR, we treat PDIs as perturbations to an otherwise homogeneous 
pump spectrum, adopting the eikonal ansatz
$(\bar{n}, \bar{u}, \bar{\phi}) = \mathrm{Re}\,(n, u, \phi)e^{i\Theta}$ with
$\Theta = \boldsymbol{K}\cdot\boldsymbol{x} - \Omega t$, where $\boldsymbol{K}$ is the
decay mode wavevector, 
\begin{equation}
    \boldsymbol{W}(t,\boldsymbol{x},\boldsymbol{k}) =(2\pi)^3\big[ \boldsymbol{F}(\boldsymbol{k}) 
    + \left(\boldsymbol{f}(\boldsymbol{k})e^{i\Theta}\right)_H\big]
\end{equation}
for the Wigner matrix, where $\boldsymbol{F}(\boldsymbol{k})$ is the unperturbed 
pump spectrum and $\boldsymbol{f}(\boldsymbol{k})$ encodes the sideband response. 
The unperturbed Wigner matrix is related to the pump potential 
spectrum $|\tilde{\phi}(\boldsymbol{k})|^2$ via the linear polarization relations; for a unidirectional pump, $F_{\alpha\beta}(\boldsymbol{k}) = l_D^4 k^4 
(\omega(\boldsymbol{k})/k_z)^{\alpha+\beta}|\tilde{\phi}(\boldsymbol{k})|^2$ (SM).
Linearizing Eq.~\eqref{eq:WME} around $\boldsymbol{F}$ and inverting for $f_{\alpha\beta}$ in terms of ($n$, $u$, $\phi$) closes the slow-field equations,  yielding the NLDR:
\begin{equation}\label{eq:nldr}
    1 + \chi_i(\Omega, \boldsymbol{K}) + \chi_{e,\mathrm{eff}}(\Omega, \boldsymbol{K}) = 0,
\end{equation}
where the effective electron susceptibility is

\begin{multline}
\chi_{e,\mathrm{eff}}(\Omega,\boldsymbol{K}) = -\frac{1}{l_D^2 K^2} \times \\[4pt]
\frac{I_{n\phi}(\Omega - I_{uu}) + (K_z + I_{nu})(v_T^2 K_z + I_{u\phi})}
{(\Omega-I_{nn})(\Omega-I_{uu}) - (K_z+I_{nu})(v_T^2K_z+I_{un})}.
\end{multline}
The coupling integrals $I_{ab}(\Omega,\boldsymbol{K})$ are resonant spectral moments of the pump:
\begin{equation}\label{eq:coupling}
    I_{ab}(\Omega, \boldsymbol{K}) = \int \frac{d\boldsymbol{k}}{(2\pi)^3} 
    |\tilde{\phi}(\boldsymbol{k})|^2 
    \frac{\mathcal{C}_{ab}(\boldsymbol{k}, \boldsymbol{K})}
    {(\Omega'- \omega')(\Omega'+ \omega')},
\end{equation}
with $\omega = \omega(\boldsymbol{k})$, $\omega' = \omega(\boldsymbol{k} + \boldsymbol{K})$, $\Omega'=\Omega+\omega$. The index structure reflects the physical role of each decay field in driving the source terms--for instance, $I_{un}$ encodes the contribution of the slow density to the ponderomotive force through pump modulation. The resonant denominator selects for pump modes satisfying the three-wave matching condition, and the coefficients $\mathcal{C}_{ab}$ encode the coupling geometry; the coupling is strongest, for instance, in the perpendicular coupling limit $\theta\to\pi/2$, where $\theta$ is the angle between the perpendicular (to $\bm{B}_0$) wavevectors of the pump $\bm{k}_{0\perp}$ and resonant sideband $(\bm{K}-\bm{k}_0)_\perp$. Explicit expressions for $\mathcal{C}_{ab}$ and $I_{ab}$ are given in the SM.
For a monochromatic pump $|\tilde{\phi}(\boldsymbol{k})|^2 \propto \delta(\boldsymbol{k} - \boldsymbol{k}_0)$, 
Eq.~\eqref{eq:nldr} is found to agree with the standard monochromatic NLDR of 
Ref.~\cite{Liu1986}. 

Equations~\eqref{eq:dens}--\eqref{eq:WME} and the nonlinear dispersion relation, Eqs.~\eqref{eq:nldr}--\eqref{eq:coupling}, together constitute the main results:
the former providing a general framework for the coupled evolution of the decay fields and pump phase-space statistics, and the latter generalizing the standard monochromatic NLDR to arbitrary broadband pump spectra $|\tilde{\phi}(\boldsymbol{k})|^2$.

To illustrate the effects of finite pump bandwidth, we solve 
Eq.~\eqref{eq:nldr} for a pump spectrum given by an anisotropic Gaussian centered 
at $\pm\bm{k}_0$ broadened along $k_z$ with width $\Delta k_{0z}$ 
while holding the perpendicular spectrum fixed at $\bm{k}_{\perp 0}$, $|\tilde{\phi}(\boldsymbol{k})|^2=2\pi^3[G(\bm{k}-\bm{k}_0)+G(\bm{k}+\bm{k}_0)]|\phi_0|^2$ where $G(\bm{k})=\delta(\bm{k}_\perp)\exp(-k_z^2/\Delta k_{0z}^2)/(\sqrt{2\pi}\Delta k_{0z})$. Because each pump component satisfies the linear dispersion relation, this parallel broadening corresponds to a relative frequency spread $\Delta\omega_0/\omega_0 \approx \Delta k_{0z}/k_{0z}$. We adopt parameters matched to the low-field-side (LFS) last closed flux surface conditions (LCFS) during an LH PDI event on Alcator C-Mod that occurred near the launcher, where the local plasma and pump wave parameters are well constrained \cite{Baek2014}. The pump amplitude $\phi_0$ is estimated from a WKB approximation \cite{Takase1983} using the scrape-off layer density profiles and input power of Ref.~\cite{Baek2014}, and the decay mode parallel wavenumber is taken as $cK_z/\omega_0 = 3$, consistent with the observed sideband peak 
frequencies. Figure~\ref{fig:spectrum} shows the maximum growth rate 
envelope $\Gamma_{max}(\Omega_R)$---the largest growth rate among all solutions at a given real 
frequency $\Omega_R$---for increasing $\Delta k_{0z}/k_{0z}$, in the perpendicular (maximum)
coupling limit $\theta = \pi/2$.

In the monochromatic limit ($\Delta k_{0z}/k_{0z} \to 0$), NRQMs fill the continuum between 
cyclotron harmonics, producing a broadband growth rate spectrum with no clear resonance 
structure. As the bandwidth increases, growth rates are suppressed across the spectrum, 
as is generically expected for finite-coherence pumps \cite{Thomson1974}. Remarkably, however, this suppression drives a fundamental reorganization of the decay spectrum: the inter-harmonic NRQM continuum is quenched far more rapidly than the peaks near cyclotron harmonics, and at sufficiently large bandwidth the spectrum collapses into a discrete set of peaks at $\Omega_R \approx n\omega_{ci}$, in qualitative agreement with the Alcator C-Mod 
observations~\cite{Baek2014}. Clear peak separation is already achieved at 
bandwidths as small as $\Delta k_{0z}/k_{0z}\approx\Delta\omega_0/\omega_0 \sim 10^{-3}$--$10^{-2}$.

This reorganization reflects a structural change in the solution set of 
Eq.~\eqref{eq:nldr}, rather than a gradual suppression of the existing NRQM solutions. Figure~\ref{fig:reorg} makes this explicit. As $\Delta k_{0z}/k_{0z}$ increases, a new branch of solutions emerges within each inter-harmonic interval, initially with small growth rate [Fig.~\ref{fig:reorg}(a)]. This branch grows in amplitude and undergoes an avoided crossing with the existing NRQM branch, after 
which the inter-harmonic solutions are eliminated entirely and only ICQMs survive. Panel~(b) shows the consequence directly: the real frequencies, initially distributed continuously across inter-harmonic values, coalesce into discrete bands localized near $\Omega_R \approx n\omega_{ci}$ as the bandwidth increases.

\begin{figure}[t]
    \includegraphics[width=\columnwidth]{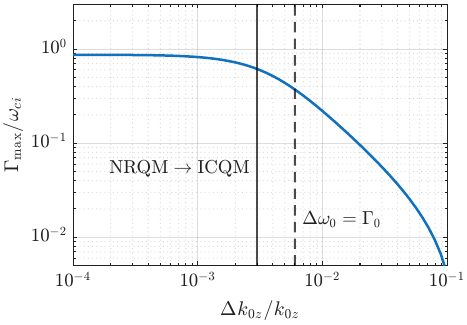}
    \caption{Maximum growth rate $\Gamma_{\max}/\omega_{ci}$ as a function of 
    parallel pump bandwidth $\Delta k_{0z}/k_{0z}$, for the same parameters as Figs.~\ref{fig:spectrum} 
    and~\ref{fig:reorg}. The solid vertical line marks the bandwidth at which the frequency gaps between all ICQM bands are fully established; the dashed vertical line marks the bandwidth at which the frequency spread equals the maximum monochromatic growth rate, $\Delta\omega_0= \Gamma_0$.}
    \label{fig:gmax}
\end{figure}

In contrast with the elimination of NRQMs, which occurs abruptly as the solution branches reconnect and the continuum solutions cease to exist, the surviving ICQMs are suppressed only gradually [Fig.~\ref{fig:gmax}]. They become effectively stabilized once the pump frequency bandwidth exceeds the monochromatic growth rate, $\Delta\omega_0\sim \Gamma_0$, beyond which the pump decorrelates before the sideband can grow — consistent with earlier broadband PDI studies \cite{Thomson1974}. As marked in Fig.~\ref{fig:gmax}, the reorganization completes at the very onset of this gradual suppression, so that the ICQM spectrum is established with little sacrifice in overall growth rate.

A full account of the physical mechanism underlying this selectivity — why one quasimode type survives finite pump coherence while the other does not — is beyond the scope of this Letter. We note, however, that NRQMs and ICQMs represent qualitatively distinct quasimode responses — NRQMs are nonresonant with $\epsilon_R \gg \epsilon_I$, while ICQMs are dissipation-dominated with $\epsilon_I \gg \epsilon_R$ \cite{Porkolab1977,Takase1983}. This distinction plausibly governs the response to a partially incoherent pump: the reactive continuum is washed out, while the dissipative peaks, anchored by ion cyclotron damping, remain comparatively robust.

Our aim here has been to isolate the qualitative effect of finite bandwidth, rather than to directly model the experimental pump spectrum, whose relation to our treatment is nontrivial. For simplicity we have considered only parallel broadening, for which frequency
and wavenumber bandwidth are directly proportional, $\Delta\omega_0/\omega_0\approx\Delta k_{z0}/k_{z0}$. The experimental spectrum does not follow this relation: the $\Delta k_z/k_{z0}$ launched by the antenna spans a substantially wider range than the values considered here \cite{Brambilla1976}, but the associated $k_{\perp 0}$ adjust to keep the frequencies near $\omega_0$, leaving a frequency spread far smaller than parallel broadening alone would produce. Edge turbulence adds further uncertainty. Nonetheless, because the NLDR~\eqref{eq:nldr} accepts an arbitrary pump spectrum, more realistic models of the experimental spectrum can be incorporated directly; we leave this to future work. The hybrid framework developed here---a fluid treatment of the nonlinear coupling combined with linear kinetic effects---may prove useful for other parametric instabilities with a similar separation of scales, such as those driven by ionospheric heating~\cite{Kuo2015}, where pump broadening is likewise expected but
has yet to be treated.

In summary, we have developed the first broadband theory of lower hybrid parametric decay and shown that finite pump coherence qualitatively reshapes the decay spectrum, collapsing the nonresonant continuum into discrete peaks at the ion cyclotron harmonics. This spectral reorganization is a novel broadband effect, distinct from the gradual suppression usually associated with finite pump coherence: it sets in at modest bandwidth, while the growth rates remain large. The resulting spectra are consistent with near-onset observations on Alcator C-Mod, which monochromatic theory fails to reproduce. Pump coherence is therefore not a mere quantitative correction to PDI growth rates but a qualitative control on the decay spectrum. More broadly, these results suggest that pump coherence may qualitatively determine which decay channels are accessible in settings beyond LHCD — wherever broadband waves drive parametric decay in the presence of kinetic resonances.

\begin{acknowledgments}
This work was supported by the U.S. Department of Energy under Awards No.~DE-‐FG02‐91ER54109, No.~DE-SC0024369, and
No.~DE-SC0014264.
\end{acknowledgments}
\bibliography{apssamp}

\end{document}


\begin{center}
  {\large\textbf{Supplemental Material}}\\[4pt]
  {\normalsize Parametric Decay of Broadband Lower Hybrid Waves}\\[2pt]
\end{center}
\vspace{1em}
\section{Derivation of main equations}
\subsection{Base model}
The base model is a drift-reduced isothermal electron fluid with linearized kinetic ions, with density normalized to the equilibrium value $n_0$ and electrostatic potential normalized as $\phi \doteq q_e\varphi/T_e$. The governing equations are
\begin{align}
  \partial_t n + \nabla\cdot(n\bm{v}) &= 0, \label{eq:cont}\\
  {[\partial_t + (\bm{v}\cdot\nabla)]} u &= -\vT^2\,\partial_z(\phi + \ln n), \\
  -\lD^2\,\nabla\cdot(1+\hat{\chi}_i)\cdot\nabla\phi &= n\label{eq:pois},
\end{align}
with $\bm{v} = \bm{v}_E + u\hat{z}$, where $\bm{v}_E = (\vT^2/\wce)\,\hat{z}\times\nabla\phi$ 
is the $E\times B$ drift and $u$ is the parallel electron velocity, $\vT = \sqrt{T_e/m_e}$, $\lD = \vT/\wpe$, $\wce = q_eB_0/m_ec$, and $\hat{\chi}_i$ the standard kinetic ion susceptibility for a magnetized Maxwellian plasma~[Stix].
\subsection{Mean and fluctuating fields}
We then consider perturbations around the equilibrium on two separated timescales,
\begin{equation}
  n = 1 + \bar{n} + \tilde{n},\quad
  u = \bar{u} + \tilde{u},\quad
  \phi = \bar{\phi} + \tilde{\phi}, \label{eq:decomp}
\end{equation}
where the averages satisfy
\begin{equation}
  \langle n \rangle = \bar{n},\quad
  \langle u \rangle = \bar{u},\quad
  \langle \phi \rangle = \bar{\phi},\quad
  \langle \tilde{n} \rangle = \langle \tilde{u} \rangle = \langle \tilde{\phi} \rangle = 0,
\end{equation}
and $\langle\cdots\rangle$ denotes averaging over timescales long compared to $\omega_0^{-1}$ but short compared to $\Omega^{-1}$. The mean fields $(\bar{n},\bar{u},\bar{\phi})$ describe the slow decay mode; the fluctuating fields $(\tilde{n},\tilde{u},\tilde{\phi})$ describe the pump and sidebands.

The mean-field equations, obtained by averaging Eqs.~\eqref{eq:cont}--\eqref{eq:pois} and retaining fluctuation-bilinear source terms while dropping products of mean fields, are
\begin{align}
  \partial_t\bar{n} + \partial_z\bar{u} &= -\langle\nabla\cdot(\tilde{n}\,\tilde{\bm{v}})\rangle, \label{eq:mean_n} \\
  \partial_t\bar{u} + \vT^2\,\partial_z(\bar{n}+\bar{\phi})
    &= -\langle(\tilde{\bm{v}}\cdot\nabla)\tilde{u}\rangle, \label{eq:mean_u} \\
  -\lD^2\,\nabla\cdot(1+\hat{\chi}_i)\cdot\nabla\bar{\phi} &= \bar{n}. \label{eq:mean_phi}
\end{align}
The fluctuating equations, under the quasilinear approximation (equivalent to the 
standard three-wave truncation), are
\begin{align}
  \partial_t\tilde{n} + \partial_z\tilde{u}
    &= -\nabla\cdot(\bar{n}\,\tilde{\bm{v}} + \tilde{n}\,\bar{\bm{v}}), \label{eq:fluc_n} \\
  \partial_t\tilde{u} + \vT^2\,\partial_z(\tilde{n}+\tilde{\phi})
    &= -(\bar{\bm{v}}\cdot\nabla)\tilde{u} - (\tilde{\bm{v}}\cdot\nabla)\bar{u}, \label{eq:fluc_u} \\
  -\lD^2\,\nabla^2\tilde{\phi} &= \tilde{n}, \label{eq:fluc_phi}
\end{align}
where the ion response is negligible at the fast timescale $\omega \sim \omega_0$ 
and $\tilde{\phi}$ is determined directly from $\tilde{n}$ via Eq.~\eqref{eq:fluc_phi}. 
These equations are linear in the fluctuating fields, but inhomogeneous due to 
coupling to the mean fields $(\bar{n}, \bar{u}, \bar{\phi})$.
\subsection{Wigner--Moyal formulation}

Equations~\eqref{eq:fluc_n}--\eqref{eq:fluc_phi} can be recast as a 
vector Schr\"odinger equation,
\begin{equation}
  i\partial_t|\bm{\psi}\rangle = \hat{\bm{H}}|\bm{\psi}\rangle, \label{eq:schrod}
\end{equation}
where $\langle\bm{x}|\psi_0\rangle = \tilde{n}$ and $\langle\bm{x}|\psi_1\rangle = \tilde{u}$, with operator
\begin{equation}
  \hat{\bm{H}} = \begin{pmatrix}
    \hat{\kvec}\cdot\bar{\bm{v}}(\hat{\bm{x}}) + \dfrac{\wpe^2}{\wce}\nabla_\perp \bar{n}(\hat{\bm{x}})\cdot\dfrac{\hat{z}\times\hat{\kvec}}{\hat{k}^2}
    & \quad\hat{k}_z\bigl(1+\bar{n}(\hat{\bm{x}})\bigr) \\[10pt]
    \bigl(\vT^2+\dfrac{\wpe^2}{\hat{k}^2}\bigr)\hat{k}_z
      + \dfrac{\wpe^2}{\wce}\nabla_\perp \bar{u}(\hat{\bm{x}})\cdot\dfrac{\hat{z}\times\hat{\kvec}}{\hat{k}^2}
    & \quad\bar{\bm{v}}(\hat{\bm{x}})\cdot\hat{\kvec} - i\partial_z \bar{u}(\hat{\bm{x}})
   \end{pmatrix}. \label{eq:Hop}
\end{equation}
The averaged Wigner matrix,
\begin{equation}
  W_{\alpha\beta}(t,\bm{x},\kvec)
  = \int d\bm{s}\,e^{-i\kvec\cdot\bm{s}}
    \langle\psi_\alpha(t,\bm{x}+\bm{s}/2)\,\psi_\beta^*(t,\bm{x}-\bm{s}/2)\rangle, \label{eq:Wdef}
\end{equation}
then satisfies the Wigner--Moyal equation (WME):
\begin{equation}
  i\partial_t \bm{W} = \bm{H}\star \bm{W} - \bm{W}\star \bm{H}^\dagger, \label{eq:WME}
\end{equation}
where $\bm{H}(t,\bm{x},\kvec)$ is the Weyl symbol of $\hat{\bm{H}}$:
\begin{equation}
  \bm{H} = \begin{pmatrix}
    k_i\star \bar{v}_i(\bm{x}) - \dfrac{\wpe^2}{\wce}\,\varepsilon_{ijz}\,\partial_i \bar{n}(\bm{x})\star\dfrac{k_j}{k^2}
    & k_z\star(1+\bar{n}(\bm{x})) \\[10pt]
    \bigl(\vT^2+\dfrac{\wpe^2}{k^2}\bigr)k_z
      - \dfrac{\wpe^2}{\wce}\,\varepsilon_{ijz}\,\partial_i \bar{u}(\bm{x})\star\dfrac{k_j}{k^2}
    & \bar{v}_i(\bm{x})\star k_i - i\partial_z \bar{u}(\bm{x})
  \end{pmatrix}, \label{eq:Hsymbol}
\end{equation}
and $\star$ denotes the Moyal product $A \star B \doteq A\,e^{i\hat{\mathcal{L}}/2}B$, with 
$\hat{\mathcal{L}} = \overleftarrow{\partial}_{\bm{x}}\cdot\overrightarrow{\partial}_{\bm{k}} 
- \overleftarrow{\partial}_{\bm{k}}\cdot\overrightarrow{\partial}_{\bm{x}}$.
It is natural to decompose $\bm{H} = \bm{H}_0 + \bm{h}$, where
\begin{equation}
  \bm{H}_0 = \begin{pmatrix} 0 & k_z \\[4pt]
  \bigl(\vT^2+\dfrac{\wpe^2}{k^2}\bigr)k_z & 0 \end{pmatrix} \label{eq:H0}
\end{equation}
encodes the free wave dynamics (its eigenvalues $\pm\omega(\kvec)$, with 
$\omega^2(\kvec) = (\vT^2+\wpe^2/k^2)k_z^2$, recover the electrostatic electron 
plasma wave dispersion relation), while
\begin{equation}
  \bm{h} = \begin{pmatrix}
    k_i\star \bar{v}_i(\bm{x}) - \dfrac{\wpe^2}{\wce}\,\varepsilon_{ijz}\,\partial_i \bar{n}(\bm{x})\star\dfrac{k_j}{k^2}
    & k_z\star \bar{n}(\bm{x}) \\[10pt]
    -\dfrac{\wpe^2}{\wce}\,\varepsilon_{ijz}\,\partial_i \bar{u}(\bm{x})\star\dfrac{k_j}{k^2}
    & \bar{v}_i(\bm{x})\star k_i - i\partial_z \bar{u}(\bm{x})
  \end{pmatrix} \label{eq:hpert}
\end{equation}
encodes all coupling to the decay fields.
\subsection{Source terms}
The source terms $S_n[\bm{W}]$ and $S_u[\bm{W}]$ appearing in 
Eqs.~\eqref{eq:dec_n}--\eqref{eq:dec_u} are the averages of bilinear 
fluctuating quantities, which can be expressed in terms of the Wigner 
matrix via standard identities of the Wigner--Weyl formalism:
\begin{align}
  S_n[\bm{W}] \doteq -\langle\nabla\cdot(\tilde{n}\tilde{\bm{v}})\rangle
  &= \int\!\frac{d\kvec}{(2\pi)^3}\!\left[
    i\bigl(W_{01}\star k_z - k_z\star W_{01}\bigr)
    + \frac{\wpe^2}{\wce}\,\varepsilon_{ijz}\,k_i\star W_{00}\star\frac{k_j}{k^2}
  \right]\!, \label{eq:source_n} \\
  S_u[\bm{W}] \doteq -\langle(\tilde{\bm{v}}\cdot\nabla)\tilde{u}\rangle
  &= \int\!\frac{d\kvec}{(2\pi)^3}\!\left[
    i\,W_{11}\star k_z
    + \frac{\wpe^2}{\wce}\,\varepsilon_{ijz}\,\frac{k_i}{k^2}\star W_{10}\star k_j
  \right]\!. \label{eq:source_u}
\end{align}

\subsection{Summary of coupled equations}

The coupled system of equations for the decay fields and pump spectrum is as follows. 
For the decay fields:
\begin{align}
  \partial_t\bar{n} + \partial_z\bar{u} &= S_n[\bm{W}], \label{eq:dec_n} \\
  \partial_t\bar{u} + \vT^2\,\partial_z(\bar{n}+\bar{\phi}) &= S_u[\bm{W}], \label{eq:dec_u} \\
  -\lD^2\,\nabla\cdot(1+\hat{\chi}_i)\cdot\nabla\bar{\phi} &= \bar{n}, \label{eq:dec_phi}
\end{align}
where
\begin{align}
  S_n[\bm{W}] &= \int\!\frac{d\kvec}{(2\pi)^3}\!\left[
    i\bigl(W_{01}\star k_z - k_z\star W_{01}\bigr)
    + \frac{\wpe^2}{\wce}\,\varepsilon_{ijz}\,k_i\star W_{00}\star\frac{k_j}{k^2}
  \right]\!, \\
  S_u[\bm{W}] &= \int\!\frac{d\kvec}{(2\pi)^3}\!\left[
    i\,W_{11}\star k_z
    + \frac{\wpe^2}{\wce}\,\varepsilon_{ijz}\,\frac{k_i}{k^2}\star W_{10}\star k_j
  \right]\!.
\end{align}
For the pump spectrum:
\begin{equation}
  i\partial_t \bm{W} = \bm{H}\star \bm{W} - \bm{W}\star \bm{H}^\dagger, \label{eq:WME2}
\end{equation}
with $\bm{H} = \bm{H}_0 + \bm{h}$, where
\begin{equation}
  \bm{H}_0 = \begin{pmatrix} 0 & k_z \\[4pt]
  \bigl(\vT^2+\dfrac{\wpe^2}{k^2}\bigr)k_z & 0 \end{pmatrix},
\end{equation}
and
\begin{equation}
  \bm{h} = \begin{pmatrix}
    k_i\star \bar{v}_i(\bm{x}) - \dfrac{\wpe^2}{\wce}\,\varepsilon_{ijz}\,\partial_i \bar{n}(\bm{x})\star\dfrac{k_j}{k^2}
    & k_z\star \bar{n}(\bm{x}) \\[10pt]
    -\dfrac{\wpe^2}{\wce}\,\varepsilon_{ijz}\,\partial_i \bar{u}(\bm{x})\star\dfrac{k_j}{k^2}
    & \bar{v}_i(\bm{x})\star k_i - i\partial_z \bar{u}(\bm{x})
  \end{pmatrix}. \label{eq:H_summary}
\end{equation}

\section{Derivation of the nonlinear dispersion relation}
\subsection{PDI formulation and equations for the Fourier coefficients}

We treat PDIs as a small perturbation to an otherwise homogeneous pump spectrum,
writing $\matr{W} = (2\pi)^3[\matr{F}(\vec{k}) + \matr{f}(t,\vec{x},\vec{k})]$ 
and $\matr{H} = \matr{H}_0(\vec{k}) + \matr{h}(t,\vec{x},\vec{k})$,
and adopting the standard eikonal ansatz
\begin{align}
  &\bar{n} = \mathrm{Re}\,\fn\,\ee^{\ii\Theta},\quad
   \bar{u} = \mathrm{Re}\,\fu\,\ee^{\ii\Theta},\quad
   \bar{\phi} = \mathrm{Re}\,\fphi\,\ee^{\ii\Theta}, \notag\\
  &\matr{f} = \bigl(\matr{\ff}(\vec{k})\,\ee^{\ii\Theta}\bigr)_H,\quad
   \matr{h} = \tfrac{1}{2}\!\left[\matr{\fh}_{(+)}(\vec{k})\,\ee^{\ii\Theta} 
   + \matr{\fh}_{(-)}^\dagger(\vec{k})\,\ee^{-\ii\Theta^*}\right]\!,
   \label{eq:eikonal}
\end{align}
where $\Theta \doteq \vec{K}\cdot\vec{x} - \Omega t$, and the perturbed Hamiltonian 
matrices $\matr{\fh}^{(\pm)}$ are
\begin{widetext}
\begin{gather}
  \matr{\fh}^{(+)} = \begin{pmatrix}
    k_{+z}\,\fu + \ii\,\dfrac{(\vec{k}\times\vec{K})_z}{\omega_{ce}}
      \!\left(\dfrac{\omega_{pe}^2}{k_-^2}\,\fn - v_T^2\,\fphi\right)
    & k_{+z}\,\fn \\[10pt]
    \ii\,\dfrac{(\vec{k}\times\vec{K})_z}{\omega_{ce}}\dfrac{\omega_{pe}^2}{k_-^2}\,\fu
    & k_{+z}\,\fu - \ii\,\dfrac{(\vec{k}\times\vec{K})_z}{\omega_{ce}}v_T^2\,\fphi
  \end{pmatrix}, \label{eq:hplus} \\[6pt]
  \matr{\fh}^{(-)} = \begin{pmatrix}
    k_{-z}\,\fu + \ii\,\dfrac{(\vec{k}\times\vec{K})_z}{\omega_{ce}}
      \!\left(\dfrac{\omega_{pe}^2}{k_+^2}\,\fn - v_T^2\,\fphi\right)
    & \ii\,\dfrac{(\vec{k}\times\vec{K})_z}{\omega_{ce}}\dfrac{\omega_{pe}^2}{k_+^2}\,\fu \\[10pt]
    k_{-z}\,\fn
    & k_{-z}\,\fu - \ii\,\dfrac{(\vec{k}\times\vec{K})_z}{\omega_{ce}}v_T^2\,\fphi
  \end{pmatrix}. \label{eq:hminus}
\end{gather}
\end{widetext}
with $\vec{k}_\pm \doteq \vec{k} \pm \vec{K}/2$. 

The decay field equations in Fourier space are
\begin{subequations}
\begin{align}
  \Omega \fn - K_z \fu
  &= \int\!\dd\vec{k}\left[K_z \ff_{01}
    - \ii\frac{\omega_{pe}^2}{\omega_{ce}}\frac{(\vec{k}\times\vec{K})_z}{(k-K/2)^2}\,\ff_{00}\right]
  \doteq S_\fn, \label{eq:decay_n} \\
  \Omega \fu - v_T^2 K_z(\fphi + \fn)
  &= \int\!\dd\vec{k}\left[\frac{K_z}{2}\,\ff_{11}
    - \ii\frac{\omega_{pe}^2}{\omega_{ce}}\frac{(\vec{k}\times\vec{K})_z}{(k-K/2)^2}\,\ff_{10}\right]
  \doteq S_\fu, \label{eq:decay_u} \\
  \ell_D^2 K^2(1+\chi_i(\Omega,\vec{K}))\fphi &= \fn. \label{eq:decay_phi}
\end{align}
\end{subequations}
Note that $-\ii S_\fu$ is the broadband generalization of the ponderomotive force. 

The Wigner perturbation $\matr{\ff}$ is related to the decay fields by the linearized WME,
\begin{equation}
  \Omega \matr{\ff} - \matr{H}_0(\vec{k}_+)\matr{\ff} + \matr{\ff}\matr{H}_0^\dagger(\vec{k}_-) 
  = \matr{\fh}_{(+)}F(\vec{k}_-) - F(\vec{k}_+)\matr{\fh}_{(-)}.
  \label{eq:linWME}
\end{equation}
\subsection{Closing the decay field equations}
To close the decay field equations, we invert the linearized WME~\eqref{eq:linWME}
to express $\matr{\ff}$ in terms of the decay field amplitudes $\fn$, $\fu$, $\fphi$.
This yields
\begin{equation}
  \begin{pmatrix}\ff_{00}\\\ff_{10}\\\ff_{01}\\\ff_{11}\end{pmatrix}
  =
  \begin{pmatrix}
    A & k_{+z}B_+ & k_{-z}B_- & C \\
    v_{p+}^2 k_{+z}B_+ & A & v_{p+}^2 C & k_{-z}B_- \\
    v_{p-}^2 k_{-z}B_- & v_{p-}^2 C & A & k_{+z}B_+ \\
    v_{p+}^2 v_{p-}^2 C & v_{p-}^2 k_{-z}B_- & v_{p+}^2 k_{+z}B_+ & A
  \end{pmatrix}
  \begin{pmatrix}R_{00}\\R_{10}\\R_{01}\\R_{11}\end{pmatrix}, \label{eq:finv}
\end{equation}
where $R_{\alpha\beta} \doteq [\matr{\fh}_{(+)}F(\vec{k}_-) - F(\vec{k}_+)\matr{\fh}_{(-)}]_{\alpha\beta}$,
\begin{subequations}
\begin{align}
  A(\vec{k}) &\doteq \bigl[\Omega^3 - \Omega\bigl(\omega_+^2 + \omega_-^2\bigr)\bigr]/D(\vec{k}), \\
  B_\pm(\vec{k}) &\doteq \bigl[\pm\Omega^2 - \omega_+^2 + \omega_-^2\bigr]/D(\vec{k}), \\
  C(\vec{k})   &\doteq -2\Omega k_{+z}k_{-z}/D(\vec{k}), \\
  D(\vec{k})   &\doteq \Omega^4 - 2\Omega^2\bigl(\omega_+^2+\omega_-^2\bigr)
                + \bigl(\omega_+^2-\omega_-^2\bigr)^2,
\end{align}
\end{subequations}
with $v_p^2(\vec{k}) \doteq v_T^2 + \omega_{pe}^2/k^2$, $\omega^2(\vec{k}) \doteq v_p^2(\vec{k})\,k_z^2$,
and we use the shorthands $v_{p\pm}^2 \doteq v_p^2(\vec{k}_\pm)$ and $\omega_\pm^2 \doteq \omega(\vec{k}_\pm)^2$.

Substituting the inversion~\eqref{eq:finv} into the source term 
integrals~\eqref{eq:decay_n}--\eqref{eq:decay_u} and shifting and flipping the integration 
variables appropriately, the source terms take the form
\begin{subequations}\label{eq:fouriersourceterms}
\begin{gather}
  S_\fn = \int\!\dd\vec{k}\,F_{00}(a_\fn\fh_{00}+b_\fn\fh_{10})
        + F_{10}(a_\fn\fh_{01}+b_\fn\fh_{11})
        + F_{01}(c_\fn\fh_{00}+d_\fn\fh_{10})
        + F_{11}(c_\fn\fh_{01}+d_\fn\fh_{11}), \\
  S_\fu = \int\!\dd\vec{k}\,F_{00}(a_\fu\fh_{00}+b_\fu\fh_{10})
        + F_{10}(a_\fu\fh_{01}+b_\fu\fh_{11})
        + F_{01}(c_\fu\fh_{00}+d_\fu\fh_{10})
        + F_{11}(c_\fu\fh_{01}+d_\fu\fh_{11}),
\end{gather}
\end{subequations}
where the shifted perturbation Hamiltonian $\matr{\fh}\doteq \matr{\fh}_{(+)}(\vec{k}_+) = 
-\matr{\fh}_{(-)}^\intercal(-\vec{k}_+)$ is given by
\begin{equation}
  \matr{\fh} = \begin{pmatrix}
    k_z'\fu + \ii\dfrac{(\vec{k}\times\vec{K})_z}{\omega_{ce}}
      \!\left(\dfrac{\omega_{pe}^2}{k^2}\fn - v_T^2\fphi\right)
    & k_z'\fn \\[10pt]
    \ii\dfrac{(\vec{k}\times\vec{K})_z}{\omega_{ce}}\dfrac{\omega_{pe}^2}{k^2}\fu
    & k_z'\fu - \ii\dfrac{(\vec{k}\times\vec{K})_z}{\omega_{ce}}v_T^2\fphi
  \end{pmatrix}, \label{eq:h_shifted}
\end{equation}
and the coupling coefficients are
\begin{subequations}
\begin{gather}
  a_\fn \doteq K_z\!\left(\frac{\omega^2}{k_z}\beta_-+\frac{\omega'^2}{k_z'}\beta_+\right)
    +\ii\frac{(\vec{k}\times\vec{K})_z}{\omega_{ce}}(v_p'^2-v_p^2)\alpha, \\
  b_\fn \doteq K_z(\alpha+v_p^2\gamma)
    +\ii\frac{(\vec{k}\times\vec{K})_z}{\omega_{ce}}(v_p'^2-v_p^2)k_z'\beta_+, \\
  c_\fn \doteq K_z(\alpha+v_p'^2\gamma)
    +\ii\frac{(\vec{k}\times\vec{K})_z}{\omega_{ce}}(v_p'^2-v_p^2)k_z\beta_-, \\
  d_\fn \doteq K_z(k_z\beta_-+k_z'\beta_+)
    +\ii\frac{(\vec{k}\times\vec{K})_z}{\omega_{ce}}(v_p'^2-v_p^2)\gamma,
\end{gather}
\end{subequations}
\begin{subequations}
\begin{gather}
  a_\fu \doteq K_zv_p^2v_p'^2\gamma
    +\ii\frac{(\vec{k}\times\vec{K})_z}{\omega_{ce}}\!\left(
    \frac{\omega_{pe}^2}{k'^2}\frac{\omega^2}{k_z}\beta_-
    -\frac{\omega_{pe}^2}{k^2}\frac{\omega'^2}{k_z'}\beta_+\right), \\
  b_\fu \doteq K_zv_p^2k_z\beta_-
    +\ii\frac{(\vec{k}\times\vec{K})_z}{\omega_{ce}}\!\left(
    \frac{\omega_{pe}^2}{k'^2}v_p^2\gamma-\frac{\omega_{pe}^2}{k^2}\alpha\right), \\
  c_\fu \doteq K_zv_p'^2k_z'\beta_+
    +\ii\frac{(\vec{k}\times\vec{K})_z}{\omega_{ce}}\!\left(
    \frac{\omega_{pe}^2}{k'^2}\alpha-\frac{\omega_{pe}^2}{k^2}v_p'^2\gamma\right), \\
  d_\fu \doteq K_z\alpha
    +\ii\frac{(\vec{k}\times\vec{K})_z}{\omega_{ce}}\!\left(
    \frac{\omega_{pe}^2}{k'^2}k_z'\beta_+-\frac{\omega_{pe}^2}{k^2}k_z\beta_-\right),
\end{gather}
\end{subequations}
with
\begin{subequations}
\begin{gather}
  \alpha(\vec{k}) \doteq A(\vec{k}_+) = A(-\vec{k}_+) = [\Omega^3-\Omega(\omega'^2+\omega^2)]/\Delta,\\
  \beta_+(\vec{k}) \doteq B_+(\vec{k}_+) = -B_-(-\vec{k}_+) = [\Omega^2-\omega'^2+\omega^2]/\Delta,\\
  \beta_-(\vec{k}) \doteq B_-(\vec{k}_+) = -B_+(-\vec{k}_+) = [-\Omega^2-\omega'^2+\omega^2]/\Delta,\\
  \gamma(\vec{k}) \doteq C(\vec{k}_+) = C(-\vec{k}_+) = -2\Omega k_zk_z'/\Delta,\\
  \Delta(\vec{k}) \doteq \prod_{\sigma_1\sigma_2}(\Omega+\sigma_1\omega+\sigma_2\omega'),
\end{gather}
\end{subequations}
written using the shorthands $\vec{k}'\doteq\vec{k}+\vec{K}$,
$k_z'\doteq k_z+K_z$, $\omega'\doteq\omega(\vec{k}')$, $v_p'\doteq v_p(\vec{k}')$. Here and throughout, we use the sign convention that $v_p \doteq +\sqrt{v_p^2}$ and $\omega \doteq v_p k_z$, 
so that the sign of $\omega$ is determined by that of $k_z$.

We can express the equilibrium Wigner matrix $\matr{F}$ in terms of the pump spectrum
for a forward-propagating pump,
\begin{equation}
  F_{\alpha\beta}(\kvec)
  = |\tilde{\phi}(\kvec)|^2\,(2\pi)^3\,(\lD k)^4
    v_p^{(\alpha+\beta)} \label{eq:Fab},
\end{equation}
where Eq.~\eqref{eq:Fab} follows from the linear wave relations for an on-shell pump,
under the assumption that different Fourier modes are statistically uncorrelated ---
which holds either for time-averaged statistics or in the presence of turbulent phase decorrelation in the plasma edge. This allows us to express the source terms in terms of
the decay-mode amplitudes as follows:
\begin{align}
  S_\fn &= I_{\fn\fn}\,\fn + I_{\fn\fu}\,\fu + I_{\fn\fsphi}\,\fphi, \label{eq:Sn_lin} \\
  S_\fu &= I_{\fu\fn}\,\fn + I_{\fu\fu}\,\fu + I_{\fu\fsphi}\,\fphi, \label{eq:Su_lin}
\end{align}
with coupling integrals $I_{ab}(\Omega,\Kvec)$ given by resonant moments (exact term?) over the pump spectrum
\begin{equation}\label{eq:couplingintegrals}
    I_{ab}= \int\!\frac{d\kvec}{(2\pi)^3}\,|\tilde{\phi}(\kvec)|^2\frac{\mc{C}_{ab}(\kvec)}{(\Omega'+\omega')(\Omega'-\omega')}
\end{equation}
with weighting coefficients
\begin{subequations}
\begin{align}
        \mc{C}_{(\fn/\fu)\fn}&=(\lD k)^4\left(v_p k_z' + i\frac{(\kvec\times\Kvec)_z}{\wce}\frac{\wpe^2}{k^2}\right) \mc{A}_{(\fn/\fu)},\\
        \mc{C}_{(\fn/\fu)\fu}&=(\lD k)^4\left[k_z' \mc{A}_{(\fn/\fu)}+\left(v_p k_z' + i\frac{(\kvec\times\Kvec)_z}{\wce}\frac{\wpe^2}{k^2}\right)\mc{B}_{(\fn/\fu)}\right],\\
        \mc{C}_{(\fn/\fu)\fsphi}&=(\lD k)^4
    \left(-i\frac{(\kvec\times\Kvec)_z}{\wce}\vT^2\right)\left(\mc{A}_{(\fn/\fu)}+v_p \mc{B}_{(\fn/\fu)}\right),
\end{align}
\end{subequations}
where
\begin{subequations}\label{eq:AsandBs}
    \begin{align}
        \mc{A}_\fn&=K_z\!\left(v_p\Omega' + v_p'\omega'\right)
    + i\frac{(\kvec\times\Kvec)_z}{\wce}
      \left(v_p'^2-v_p^2\right)\Omega',\\
      \mc{B}_\fn&=K_z\!\left(\Omega' + v_pk_z'\right)
    + i\frac{(\kvec\times\Kvec)_z}{\wce}
      \left(v_p'^2-v_p^2\right)k_z',\\
      \mc{A}_\fu&=K_z v_p v_p'\omega'+i\frac{(\kvec\times\Kvec)_z}{\wce}\left(
      \frac{\wpe^2}{k'^2}v_p\Omega'
      - \frac{\wpe^2}{k^2}v_p'\omega'\right),\\
    \mc{B}_\fu&=K_z v_p \Omega' +i\frac{(\kvec\times\Kvec)_z}{\wce}\left(
      \frac{\wpe^2}{k'^2}v_p k_z'
      - \frac{\wpe^2}{k^2}\Omega'\right).
    \end{align}
\end{subequations}
and we use the shorthand $\Omega'=\Omega+\omega$.

\subsection{Effective electron susceptibility and NLDR}
Substituting Eqs.~\eqref{eq:Sn_lin} and \eqref{eq:Su_lin}
into Eqs.~\eqref{eq:decay_n}--\eqref{eq:decay_u}
we can solve for $\fn$ in terms of $\fphi$ and substitute into the slow Poisson
equation [Eq. \eqref{eq:decay_phi}]. The result takes the form
\begin{equation}
  \bigl[1 + \chi_i(\Omega,\Kvec) + \chi_{e,\mathrm{eff}}(\Omega,\Kvec)\bigr]\phi = 0,
  \label{eq:NLDR_form}
\end{equation}
where the effective electron susceptibility is
\begin{equation}
  \chi_{e,\mathrm{eff}}(\Omega,\Kvec) \doteq -\frac{1}{\lD^2 K^2}
  \frac{I_{n\phi}(\Omega - I_{uu}) + (K_z + I_{nu})(v_T^2 K_z + I_{u\phi})}
  {(\Omega - I_{nn})(\Omega - I_{uu}) - (K_z + I_{nu})(v_T^2 K_z + I_{un})}.
  \label{eq:chi_eff}
\end{equation}
In the absence of the pump ($I_{ab} \to 0$), this reduces to
\begin{equation}
  \chi_{e,\mathrm{eff}} \to \frac{1}{\lD^2 K^2}\frac{1}{1-(\Omega/K_z\vT)^2},
  \label{eq:chi_e_linear}
\end{equation}
which is just the scalar linear susceptibility of the isothermal drift-reduced fluid model. The NLDR follows directly from Eq.~\eqref{eq:NLDR_form}:
\begin{equation}
  \boxed{1 + \chi_i(\Omega,\Kvec) + \chi_{e,\mathrm{eff}}(\Omega,\Kvec) = 0,}
  \label{eq:NLDR}
\end{equation}
where $\chi_{e,\mathrm{eff}}$ is defined in Eq.~\eqref{eq:chi_eff} and the coupling terms therein
are defined in Eqs.~\eqref{eq:couplingintegrals}--\eqref{eq:AsandBs}. Equation~\eqref{eq:NLDR} is the
main result [Eq.~(9) of the Letter], constituting the broadband generalization of
the standard monochromatic NLDR.